\documentclass[graybox]{svmult}

\usepackage{mathptmx}       
\usepackage{helvet}         
\usepackage{courier}        
\usepackage{type1cm}        
\usepackage{makeidx}         
\usepackage{graphicx}        
\usepackage{multicol}        
\usepackage[bottom]{footmisc}

\usepackage{natbib}		
\usepackage{url}

\makeindex             

\begin{document}

\title*{Big Data}
\author{Andreas L. Opdahl\inst{1} \and Vimala Nunavath\inst{2,1}}
\authorrunning{A.L.~Opdahl \& V.~Nunavath}
\institute{
	(1)~University of Bergen, 
	Norway, \email{Andreas.Opdahl@uib.no}
\and
    (2)~University of Agder, 
    Norway, \email{Vimala.Nunavath@uia.no}
}
%
%
\maketitle

\abstract*{The Internet of Things, crowdsourcing, social media, public authorities, and other sources generate bigger and bigger data sets. Big and open data offers many benefits for emergency management, but also pose new challenges. This chapter will review the sources of big data and their characteristics. We then discuss potential benefits of big data for emergency management along with the technological and societal challenges it poses. We review central technologies for big-data storage and processing in general, before presenting the Spark big-data engine in more detail. Finally, we review ethical and societal threats that big data pose.}

\abstract{The Internet of Things, crowdsourcing, social media, public authorities, and other sources generate bigger and bigger data sets. Big and open data offers many benefits for emergency management, but also pose new challenges. This chapter will  review the sources of big data and their characteristics. We then discuss potential benefits of big data for emergency management along with the technological and the societal challenges it poses. We review central technologies for big-data storage and processing in general, before presenting the Spark big-data engine in more detail. Finally, we review ethical and societal threats that big data pose.}

\section{What is Big Data?}
%
%

There is already much more data available on the internet than humans can meaningfully process. Spread out among this data may lie central keys to prevent and better manage emergencies. To unlock the knowledge they contain, big data sets must become effectively processable by machines and the results easily interpretable for humans~\citep{berners-lee_semantic_2001}.

Big data is a broad term with many related meanings on the technical, computing, data, and usage levels. On the \textit{technical-infrastructure level}, big data has often been used about data collections that are too large to be straightforwardly handled with traditional mainstream data-processing techniques and tools. Starting in the 2000s, big internet companies like Google, Amazon, and Facebook found they needed new ways and new software tools to store and process the enormous data collections they were amassing at the hearts of their businesses. The result was a new generation of distributed technologies we will review later in this chapter: file systems such as HDFS; grid and cloud technologies such as Amazon's Web Services (AWS); big-data stores such as Cassandra; and big-data processing engines such as Spark. 

On the \textit{computing level}, big-data computing methods also differ from past mainstream approaches. Most importantly, they need to be highly distributed because computing, networking, and storage demands can go way beyond the abilities of single computers. In consequence, computing must also be fault-tolerant because, when a distributed system consists of hundreds or thousands of computers, failing components become the norm rather than the exception. To achieve this, big-data computing must therefore be highly redundant, so that each computation is carried out and each piece of data stored in several places simultaneously, and so that computation can go on even when some of the computers, disks, or networks fail. While solutions to each of these problems have existed for a long time, big-data computing has brought them together in new techniques and tools and moved their use from the fringe into the mainstream~\citep{ghemawat_google_2003,dean_mapreduce:_2008,lakshman_cassandra:_2010,zaharia_apache_2016}.

On the \textit{data level}, big data refers to \emph{``the three Vs''}: data that has large \textit{volume}, arrives at high \textit{velocity}, and has great \textit{variety} --- consisting of both structured data, natural language, images, audio, and video~\citep{kitchin_data_2014}. While earlier mainstream approaches could deal with two of these at the expense of the third, the ability to support all three ``Vs'' at the same time is a central requirement of big data. Two more ``Vs'' are that big data must be \textit{valid} or true (\textit{veracity}) and provide \textit{value} to users. In addition, big data aims to be \textit{exhaustive}, by representing each and every relevant phenomenon rather than just a sample. It tends to be \textit{fine-grained}, by representing every available piece of information in the life-cycle of a phenomenon. It is \textit{indexical}, in that it uses, as much as possible, standard identifiers for phenomena, attributes, and events. It is \textit{relational}, in the sense that information about different phenomena are connected by use of standard identifiers. It is also easily \textit{extendible} to new types of phenomena, attributes and events and \textit{scalable} in size as more information is added. It is \textit{historical}, by representing not just current but also all previous states of phenomena. And it can be \textit{opportunistic}, in the sense that data is stored exhaustively, historically and in full detail because it can potentially generate value in the future, even if it is not yet needed today~\citep{kitchin_data_2014}. Despite including the word ``big'', \textit{data size} is less used as a central characteristic of big data. Although the big global players today manage data collections in the peta- ($10^{15}$) and exabyte ($10^{18}$) range, many smaller- and medium-sized companies find big-data techniques essential for creating value from datasets with high variety and velocity, even when the volumes are much smaller.

Finally, on the \textit{usage level}, big data refers to new data-driven ways of managing and organising private, public, and ideal enterprises. So-called \textit{data-driven projects, businesses/organisations, and societies} continuously harvest exhaustive, fine-grained data about their inner workings and environment. They process the data in real time, using machine learning techniques on historical data to continuously describe and diagnose the past and present in order to optimally predict the future and prescribe optimal actions in advance~\citep{kitchin_data_2014}.

\section{Big Data Sources for Emergency Management}

Several organisations provide online overviews of big and other data sources that can be useful in emergencies. The Humanitarian Data Exchange (HDX\footnote{\url{https://data.humdata.org}}) indexes more than 6000 crisis-relevant data sets, searchable by features, location, format, organisation, license, year, and general tags. CSV-tables is the most common data format by far. PreventionWeb\footnote{\url{http://www.preventionweb.net}} offers a portal to diaster-related datasets and sites across the globe, focussing on past disaster loss and damage, historical and synthetic hazard catalogues, socio-economic factors that can impact vulnerability and resilience on a local level, and exposure data about populations and buildings in particular locations.

General open data sources can also serve as useful references in emergencies. GeoNames\footnote{\url{http://geonames.org}} is a global database of geographical features that is searchable and browsable through a map interface. OpenStreetMap\footnote{\url{http://openstreetmap.org}} is similar, but provides more detailed information about populated areas. The Humanitarian OpenStreetMap\footnote{\url{http://www.hotosm.org/}} is an international network dedicated to humanitarian action and community development through open mapping, for example of refugee situations, volcanic eruptions, and ebola outbreaks. Wikipedia is an important source of open reference data, along with its more recent sister project Wikidata\footnote{\url{http://www.wikidata.org/}} for factual and structured data. Google Crisis Map and Person Finder\footnote{\url{http://www.google.org/\{crisismap,personfinder\}}} are examples of proprietary data resources that can be leveraged in crisis situations through well-defined interfaces.

Public authorities on the international, national, and local levels are also important providers of reference data, for example census and map data. Many authorities also maintain emergency-relevant data about buildings (kindergartens, schools, hospitals, care facilities) and statistical data about their population. More sensitive governmental data include data about critical infrastructures: water supply and sewage, electricity, communication, roads and railways, and military installations and operations. Some local governments may also maintain information about vulnerable citizens that need particular assistance. In many countries, a wide variety of emergency-related data sets are already available through portals and data hotels such as \url{http://data.europa.eu} in the EU and \url{http://data.gov} in the US. Unfortunately, many datasets provide only static, aggregated, historical statistics. There are fewer live web APIs that offer minute-to-minute information. 

There are also many closed (for-pay or restricted) data sources provided by businesses and governments. Some of them have higher quality than their open counterparts in terms of completeness, correctness, precision, timeliness, etc. Examples of private and semi-private companies that maintain potentially emergency-related information are: mapping agencies (maps, buildings, critical infrastructure); telecommunications companies (location and movement of subscribers, communication patterns, habits, pictures); transport companies (movement patterns, locations of people and vehicles); and app providers (many of the above and more). In some countries, previous government agencies have been privatised or semi-privatised in recent years, making emergency-relevant data sources less accessible. 

The Internet of Things (IoT) is another source of big data, whose importance is rapidly growing~\citep{atzori_internet_2010}. There are already many times more things connected to the internet than there are people. These things can be sensors that measure and observe, such as a thermometer or surveillance camera, or they can be actuators that change the state of physical things, such as an alarm or a traffic light. Many things on the internet, such as mobile phones, can be both sensors and actuators, and they are quickly becoming smarter. Some of them have enough processing power and storage capacity to run heavy computing tasks locally, and others run apps that collaborate tightly with software agents running on more powerful computers in the cloud. As the Internet of Things continues to grow, it will offer enormous opportunities for preventing, detecting, limiting, managing, and recovering from emergency situations, which we discuss in the following section. 

Of course, social media is another central source of big data. It is so important that we devote a separate chapter to it.
For research purposes, CrisisLex\footnote{\url{\https://www.crisislex.org/data-collections.html}} contains brief descriptions of downloadable emergency-related social-message collections from the past. In a later chapter we will also discuss emergency-related datasets that are available in semantic formats (such as RDF and OWL).

\section{Big Data Benefits and Challenges}

\subsection{Benefits}
Big-data analytics can be helpful in all phases of emergency management, from preparation, through detection and response, to recovery~\citep{castillo_big_2016}.

For \textit{preparation} purposes, big data can be used to create baseline models that describe and diagnose (explain) normal conditions, such as the normal movement of people and goods in a city; normal consumption of transport services, power and water; normal geographical and meteorological conditions; and normal physical conditions inside a building. Diagnosing these and other conditions calls for big-data analytics because they may be highly dependent on contextual factors, not only on the most obvious ones, such as the location, time-of-day, day-of-week, and part-of-year, but also on weather, holidays and their types, public health situation, state of the economy, accidents and emergencies, culture and sports events (both local and global, such as the football World Cup finals). Baseline models can be used to quickly identify deviating conditions and as input to simulating likely consequences of deviations, such as choke points or single points of infrastructural failure. Baselines can also be used for emergency preparation and training and for post-hoc analysis after the crisis or emergency. 
Baselines can be created from measurements and physical-observations from IoT devices and from social media sensors (see the chapter on social media in this book) that gauge people's moods and concerns, augmented with contextual information from private and public data sets, such as maps and information about buildings, infrastructure, population, vulnerable citizens, etc.

For \textit{detection} purposes, potential emergency situations can be identified early by continuously monitoring people and their environments. In some cases, a quick response can even prevent an unstable or crisis situation from evolving into a full-blown emergency. Useful common sensor types include: surveillance cameras, mobile phones, wearable devices, and social media messages that signal people who are dangerously crammed together at a public event; all kinds of indoor heath, humidity, temperature, and light sensors that can inform about deviations such as break-ins, fires and leakages; positioning devices in vehicles, combined with mobile phone positioning data, social media messages, traffic sensors, road cameras, and live weather data that indicate traffic accidents --- or increased risk of such; and, of course, all kinds of weather sensors combined with satellite images that give the earliest warning possible about looming deviations from normal geophysical and meteorological conditions.

For \textit{response} purposes, similar types of information can be used during an unfolding emergency both to gain (strategic) situational overviews and (tactical) actionable insights~\citep{castillo_big_2016}, for example using map-based visualisations and interfaces. Information from physical and social sensors can be corroborated with open and reference data to improve data-quality attributes such as correctness and trustworthiness. Actuators can be repurposed to actively relieve and improve the situation. For example, a traffic light or gated crossing can be taken out of normal operation and instead be used to direct traffic around an affected area. Remote-controlled drones and other vehicles can be used to collect detailed information and to disseminate medical equipment and food.  Social media and other communication technologies can be used in a similar manner (as \textit{social actuators}) to disseminate results of big-data analysis and other information back to responders, victims, and their families and friends.

For \textit{recovery} purposes, big-data technologies can be used to make full data traces of the emergency available for post-mortem analysis, both of the unfolding disaster itself, of its preconditions, of the the detection and response effort, and of their impacts. Of course, emergency-data traces, when combined with physical and social baselines and reference data also have numerous uses in the aftermath of an emergency: for rebuilding and to prevent or prepare for similar situations in the future.

\subsection{Challenges}
Properly leveraging big-data technologies for emergency management also poses many challenges. Most of all, big-data harvesting, preparation, curation, analysis, and interpretation is skill intensive process and demand for big-data competency is vastly greater than supply. Recruiting people with the right competencies, both as professional emergency workers and as volunteers, is therefore a challenging task. Global networks-of-networks are in place for recruiting, training, and coordinating volunteers in various types of emergency and environment related activities, for example in global-mapping activities or satellite-image analysis. Subnetworks of volunteers with competencies in big-data analytics are called for.

Availability of computing infrastructures is a also an issue, both at the emergency site and remotely. Cloud computing makes it feasible to create big-data infrastructures in advance. After they have been set up and tested, the cloud resources and services can be paused until an emergency occurs. Then they can be restarted quickly and easily scaled when computing demands increase. However, advance-preparing for an emergency is hard. Some of the data and processing needs that arise during an emergency will always be unexpected, and preparations must therefore focus on flexibility and responsiveness to change. 
Availability of computation and communication facilities near the emergency site can be a bigger issue. For example, local sensor networks may be damaged or disconnected due to network or power failure. Aerial-network drones and store-and-forward networking are possible ways to temporarily reinstate networking capacity in affected areas. Disseminating analysis results to the affected people is also a challenge, because most of the techniques and tools have been developed for highly-trained scientists, business people, and other skilled decision makes. Communicating the results of complex big-data analyses to emergency workers, victims, and their families and friends most likely requires different presentation techniques. 
In a crisis, criminals, terrorist groups, or hostile powers may try to exploit or destabilise the situation further. Social media, which can be used to spread disinformation (misinformation deliberately intended to deceive), has become a well-known attack vector. Cyberattacks are another possibility, for example to interfere with critical computer-controlled infrastructures, to hi-jack or cut off sensors and actuators, or to compromise the emergency-computing infrastructure itself.

Other challenges are less specific to emergency management. High-velocity data must be collected in real time from different data sources, of different types, in different modalities (audio, written text, images, structured data, etc.), using different formats and access methods (social media, the web, FTP, web APIs, etc.). Widely heterogeneous data sets must be recombined and corroborated to create richer situational overviews and facilitate more reliable actionable insights. The data, in particular from social media, must be filtered to clear large amounts of noise. Trustworthiness and other quality features (timeliness, completeness, etc.) must be assessed. 

The many ethical issues of big data will be discussed in a later section. 

\section{Big Data Techniques and Tools}

Although it builds on theories and techniques that are much older, the big-data wave gained momentum in the mid 2000s, when data-driven businesses and big internet companies like Amazon, Facebook, and Google needed new ways to store and process the rapidly growing data sets they had amassed. The result was a new generation of techniques and tools for big-data processing that includes: cloud and grid computing technologies, distributed file systems, new types of database management systems, and new distributed computing engines. Many of the ideas they build on are old, but the tools themselves are new, and big data have made their use much more widespread. 

Big data is made possible by improvements in computing power, storage capacity, and network bandwidth~\citep{kitchin_data_2014}. In addition, new sources of large data sets have emerged, such as social media, the Cloud/Internet of Things (IoT/ClouT), and ubiquitous and pervasive computing. Important enablers on the computing side are cloud and grid techniques. \textit{Grid computing} lets many loosely-coupled computers distributed across the net perform large computing tasks together. \textit{Cloud computing} leverages grids to offer highly scalable services to end-users, such as massive data storage, powerful virtual and remote servers, big-data analysis services, and much more. 

For example, Amazon Web Services (AWS\footnote{\url{https://aws.amazon.com/}}) uses large grids to offer a wide range of scalable storage, computation and analysis services in the cloud. AWS' Elastic Compute Cloud (Amazon EC2\footnote{\url{https://docs.aws.amazon.com/AWSEC2/latest/UserGuide/concepts.html}}) servers appear to their users as regular computers accessible over the net, but they are virtual: each of them may be running on only a small part of a physical machine, or it can run on a large grid of machines. Virtual servers can be quickly and easily created, started, stopped, restarted, resized, replicated, and even moved between data centres on different continents as computing demands change. A new cloud server can be instantiated in minutes, either as a clean Linux or Windows computer or by instantiating an existing Amazon Machine Image (AMI) that represents a predefined computer configuration, such as a big-data set-up designed specifically for emergency computing.
There are two common approaches to create an AMI in Linux. The first approach is to start from an existing public AMI and modify it according to the users' requirements. 
The second approach is to build a fresh installation, either on a standalone machine or on an empty file system mounted by loopback~\citep{Huang2010}.
Amazon also offers a choice between two types of storage services: Elastic Block Store (EBS) and Simple Storage Service (S3), along with management services such as checkpointing and security policies.

New file systems such as the Google File System (GFS, later renamed Colossus)~\citep{ghemawat_google_2003} and Apache's open-source HDFS (Hadoop Distributed File System\footnote{See \url{https://hadoop.apache.org/}.}) are able to handle big-data volumes by distributing data storage across many computers, possibly in the tens of thousands. On these scales, data storage (and computing) must be fault-tolerant because failing components become the norm rather than the exception. Each big-data file (possibly of tera- or petabyte size) is therefore split into blocks (e.g., 128Mb each) that are stored on different computers, called data nodes in HDFS. To achieve fault tolerance, the same block is replicated on several data nodes (\textit{sharding}). Inside a data centre, the nodes are grouped  into racks that correspond to physically co-located sub-networks of computing nodes served by the same router. HDFS attempts to shard each block across nodes that belong to different racks so that, if a rack becomes unavailable due to router failure or power outage, other copies of the blocks it stores will be available from other racks. A single name node handles client requests and keeps track of where the blocks in each file are stored. The name node thus makes HDFS appear as a single logical file store to client programs, although data is transmitted directly between clients and data nodes. Communication with clients and between HDFS nodes use TCP/IP, making it easy to include heterogeneous computing nodes in the same cluster (which is then often called a grid). To avoid making the name node a single-point of failure, other data nodes are continuously monitoring their name node and are always ready to step up should it lose connection. As with many big-data technologies, HDFS is optimised for mostly immutable files: the initial write and the subsequent reads of a file are much faster and use less resources than updating the file after it has been created.

The new distributed file systems quickly inspired similarly distributed database management systems (DDBMS), designed to accommodate big-data collections (tera- and petabytes) in a flexible way. In addition to supporting higher numbers of data items of the same type (\textit{vertical scaling} in volume), new big-database systems also emphasise supporting rapidly evolving information needs of different types (\textit{horizontal scaling} in variety). To achieve this, they deviate in several ways from traditional relational and SQL-based data models, giving them the name Not Only SQL (NOSQL) databases. MongoDB uses the JavaScript Object Model (JSON) as its data model for both storage and interaction. Google's Bigtable~\citep{chang2008bigtable} organises data as sparsely populated tables, called wide-column stores, and has inspired both Apache HBase, which runs on top of HDFS, and Apache Cassandra~\citep{lakshman_cassandra:_2010}. The latter has become one of the most widely used big-data stores. 
Initially developed to support Facebook's messaging system, Cassandra combines features from key-value pairs and wide-column stores. Like most NOSQL databases, Cassandra is designed to support data replication, resilience towards failure, and ease of adding more machines to the cluster/grid. It offers flexibility, scalability and read-orientation: like HDFS, it is optimised for writing new data once and reading it many times, whereas updating existing data is more costly. Unlike HDFS, however, all Cassandra nodes are equal: they can both store data and answer client requests. 

Like data storage, processing has become distributed and duplicated across computing nodes and centres too. The original Google MapReduce~\citep{dean_mapreduce:_2008} and Apache's open-source Hadoop\footnote{\url{https://hadoop.apache.org/}} support a massively parallel three-step model of computation called \textit{map-reduce}. In the first step (\textit{map}), each computer, called a worker (or slave) node in Hadoop, sorts the input data into local files according to a key. In the second step (\textit{shuffle}), the nodes transfer temporary files between them, so that all data items with the same key end up in the same worker node. In the third and final step (\textit{reduce}), each worker node processes the data for its keys, possibly combining its own results with results from its neighbour nodes. For example, to rapidly identify duplicates in a set of merged library catalogues, the map step sorts each item locally by title, the shuffle step moves all items with the same title to the same worker, and the reduce step decides whether pairs of items that happen to have the same title are actually duplicates. A Hadoop worker node can be the same computer that is also a data node in HDFS or database node in Cassandra, so that the node works on locally stored input data in the initial map step. Otherwise, a costly input split has to be carried out before first.

A Hadoop master node handles client requests and distributes tasks to the workers, ensuring that redundantly stored blocks are only processed by one worker each. Hence, in Hadoop, as in most other big-data processing frameworks, the tasks are moved to the data. This is a shift from conventional (``small-data'') computing, where the data was moved to the task processor. In addition, the master node is responsible for balancing tasks between workers, taking into account their varying processing capacities, and for reallocating tasks when a worker is slow or becomes unavailable. Many extensions and variations of the basic map-shuffle-reduce model have been proposed, and multiple Hadoop computations are often chained. Because it has been so successful, Hadoop's subsystem for resource management and job scheduling and monitoring has been turned into a separate framework, called YARN (Yet Another Resource Negotiator) that can be used by other big-data processing tools as well.

These tools and their successors share a model of computing that is both highly distributed and highly redundant, so that each computation is carried out and each piece of data stored in several places simultaneously. On top of node-level replication, whole data centres can be replicated (or mirrored) across continents. Data-centre replication improves service availability and can also be used as a content delivery network (CDN) that improves responsiveness and communication cost by distributing services geospatially according to user locations: requests from Chinese users can be served by a replicated data centre in Guangzhou, European requests can be routed to Frankfurt, and so on. 

In recent years, more specialised big-data technologies and tools have emerged, many of them growing out of the HDFS/YARN/Hadoop stack. Apache Hive is a data warehousing tool that offers an SQL-like query language on top of HDFS and converts the queries to Hadoop, Spark (see below), or other types of big-data jobs. Apache Kafka and Amazon Kinesis specialise in processing and storing big streams of data, which can be produced internally in an organisation or harvested externally from the social web or the Internet of Things. Google's Pregel~\citep{malewicz_pregel:_2010} and Apache Giraph\footnote{\url{http://giraph.apache.org/}} specialise in processing and storing big graph-structured data. They have been used, respectively, to drive Google's PageRank algorithm~\citep{page1999pagerank} and Facebook's social network analyses. And many very large-scale applications still use good-old SQL databases. Alongside these and many other specialised technologies, a preferred tool for general big-data processing has emerged: Apache Spark, which we will present in the next section.

\section{General Engine for Big Data Processing: Spark}

The usefulness and power of Google's MapReduce and its successors surpassed many expectations. But this first generation of big-data processing engines also had severe limitations. They were batch-oriented, with jobs that could take days because they were heavily disk-based, copying their intermediate results repeatedly to and from disks. Their computation model was also rigid, restricted to chaining jobs that were composed of minor variants of the map, shuffle, and reduce operations.

In the early 2010s researchers therefore sought to develop big-data processing tools that were more interactive; that reduced disk load by relying on in-memory data storage; and that offered a broader variety of computing primitives. The most widely used among them is Spark~\citep{zaharia_apache_2016}, which is part of Apache's big-data ecosystem. In addition to Hadoop's map, shuffle, and reduce operations and their variants, Spark offers more than 80 different high-level processing operations, which can be detailed with functions written in standard programming languages such as Python, Java, Scala, and R. There are also Spark libraries for: connecting to Cassandra and SQL databases; analysing large graphs (GraphX); processing live data streams (DStream); and machine learning (MLib). Like in Hadoop, worker nodes do most of the computing. A driver node communicates with the client and distributes tasks to workers, whereas a cluster manager (either Spark's own built-in one, YARN, or another) allocates processing capacity.

A Spark computation is organised as a dataflow (or pipeline). The dataflow can be thought of as a directed acyclic graph (DAG) where each node represents data at some stage of processing and each edge represents a processing operation. The data in each node is treated as a Resilient Distributed Dataset (RDD)~\citep{zaharia_resilient_2012}, which may contain anything from a single boolean variable to a petabyte collection of videos:
the data in an RDD can be structured like key-value pairs (maps), tables, and graphs or unstructured like text and multimedia. But all data items in the same RDD must have the same type, which is written RDD[\textit{type}]. Because of its size, an RDD is usually split into partitions that can be distributed across thousands of computing nodes, where they are stored in memory and normally not copied to disk, unless memory fills up or the user specifically requests a checkpoint to be saved. Importantly, RDDs are immutable meaning, that once they have been created by a computing operation, they never change (more about that below).

Spark operations take input from one or more nodes and may produce output to a single node. They are divided into transformations, which take one or more RDDs and generate a new RDD, and actions, which take one or more input RDDs and produce either an output (that is not an RDD) or a side effect. For example, Spark \emph{transformations} are available for mapping, filtering, reducing, sampling, and sorting RDDs in various ways, and for combining multiple RDDs using operations such as intersection, union, and join. After each transformation, the result is always a new RDD. Spark \emph{actions} are available for counting, sampling, looping through, or otherwise reducing large RDDs into outputs that are easier for humans to interpret and for simpler software tools to process. Other actions are used for their side effects, for example to cache or checkpoint an RDD or save it to a (distributed) file system. Spark evaluation is lazy, so that a transformation in a dataflow will not be executed until the RDDs they produce are needed, either directly or indirectly, as input to an action.

Spark processing is resilient because both data storage and processing is highly redundant: each part of an RDD and of a computation can be stored and processed on several cluster nodes simultaneously. Whenever a node fails during computation, and that part of the computation is duplicated, the computation can just go on. Whenever a node fails that is not redundantly processed, the Spark engine instead relies on \emph{lineage}. It automatically enforces recomputation of the necessary data from the last available checkpoints, perhaps even going back to the original inputs. Recomputation is possible exactly because the RDDs are immutable: once created they never change, so each later processing step can be recomputed safely.

To program using Spark, we can use languages like Python, Java, Scala, and R along with libraries that make Spark operations and RDDs available from inside these languages. We can also  program interactively using Spark's built-in shells for Python and Scala. The following example is written in Scala using the spark-shell, but the code in other languages would not be much different. 

To filter Twitter data with Spark in an emergency, we first import Spark's DStream library and its Twitter extensions\footnote{%
See the Apache Spark DStream and Bahir-Twitter projects. The Twitter4j library is also needed.}:

{\small\begin{verbatim}
	import org.apache.spark._
	import org.apache.spark.streaming._
	import org.apache.spark.streaming.twitter._ 
\end{verbatim}}\noindent

We also need to specify the credentials for our Twitter account and app, with lines like\footnote{%
To register and get credentials for a Twitter App, go to apps.twitter.com .}: 

{\small\begin{verbatim}
	System.setProperty( "twitter4j.oauth.consumerKey", "..." )  
	System.setProperty( "twitter4j.oauth.consumerSecret", "..." )  
	System.setProperty( "twitter4j.oauth.accessToken", "..." )  
	System.setProperty( "twitter4j.oauth.accessTokenSecret", "..." ) 
\end{verbatim}}\noindent
We are now ready to define a dataflow that harvests tweets from Spark (although nothing will be executed until we start the stream and specify an action):

{\small\begin{verbatim}
	sc.setLogLevel( "ERROR" ) 
	val ssc = new StreamingContext( sc, Seconds( 5 ))  
	val stream = TwitterUtils.createStream( ssc, None )
\end{verbatim}}\noindent

The first line states that the running Spark engine, represented by the built-in Spark Context (sc) object, should only report serious errors. The second line states that Spark should run in streaming mode, collecting a new batch of input data every five seconds. The third line states that this data should come from the Twitter account we have already specified. 

The stream variable represents the series of RDDs that are input to our pipeline, one every five seconds. Each RDD in it has the Scala type RDD[\textit{Status}], meaning that it is an RDD that contains Status objects. The Status class is defined in the Twitter4j library to represent and process a single tweet along with its (quite extensive) metadata\footnote{%
See the introduction to Tweet JSON at developer.twitter.com and the twitter4j.Status interface in the Twitter4j API, which Bahir-Twitter wraps around.}. 

We can easily add more operations to our dataflow. First, we use a map transformation to pick out the message texts from each tweet:

{\small\begin{verbatim}
	val texts = stream.map( status => status.getText )
\end{verbatim}}\noindent

This creates a new stream of RDDs of strings (type RDD[\textit{String}]), which we can loop through and output (an action):

{\small\begin{verbatim}
	texts.foreachRDD( _.foreach( text => println( text )))
\end{verbatim}}\noindent

or split (a transformation) into a stream of RDDs of single words (again of RDD[\textit{String}]):

{\small\begin{verbatim}
	val words = texts.flatMap( text => text.split( “ ” ))
\end{verbatim}}\noindent

which we then filter (another transformation) for hashtags:

{\small\begin{verbatim}
	val hashtags = words.filter( _.startsWith( "#" ))
\end{verbatim}}\noindent

We can easily add further operations that may extend, split or merge processing paths. When we are finished, we can start the flow of tweets and inspect the outputs:

{\small\begin{verbatim}
	ssc.start
\end{verbatim}}\noindent

In the example, this will start extracting message texts from live Twitter messages and output them to the console. But it will not start splitting texts into words and filter out the hashtags, because Spark evaluation is lazy and we have not yet used the hashtags in an action. 

Finally, we stop the Spark streaming pipeline, making it clear that we want to wait until all data in the pipeline has been processed and that we do not want to close the Spark context permanently:

{\small\begin{verbatim}
	ssc.stop( stopSparkContext=false, stopGracefully=true )
\end{verbatim}}\noindent

Of course, this small example barely scratches the surface of what a powerful big-data framework like Spark can do. But it illustrates how just a few lines of Spark code is enough to process social media data in ways that are potentially useful in emergency situations. In addition to its high-level processing operations, much of the power of Spark lies in its scaling: the above example can be run both on a single computer on a filtered stream of tweets and, without modifications, on a cluster of hundreds or thousands of nodes processing a fire hose of data. Of course, the Spark configuration would have to be changed, but there are many abstract machine images for distributed Spark freely available in the cloud, and instantiating one of them only takes minutes.

\section{Ethical and Societal Issues}

The previous sections have shown that big-data analytics offer many benefits in all phases of emergency management.  New possibilities for large-scale yet precise surveillance on demand can save lives and property in emergency situations. But if the same surveillance is misused in everyday life, it can pose threats to individual privacy and to society in general. Use of big data during emergencies must therefore be carefully conducted and monitored, and a fine line must be tread between specific emergency needs and wider ethical and societal concerns~\citep{kitchin_data_2014}.

Among the ethical concerns, privacy is central. \textit{Personal information} refers to information that can be attributed to a physical person. This includes data that contains unambiguous identifiers such as personal id numbers, names, addresses, and birth dates, but also data that is sufficiently detailed to be indirectly attributable to a person. A study of mobile phone-users~\citep{de2013unique} showed that knowing only four spatio-temporal data points was enough to identify 95\% of the users uniquely, and coarser datasets did not offer much stronger anonymity. Hence, when data sets about individuals are recombined in an emergency situation, the combined data items can become attributable to individuals, even when the original data were not. Privacy concerns are accentuated when the personal information is also \textit{sensitive} because it covers: age, criminal records, ethnicity, gender, health, marital status, political opinions, race, religion, sexuality, or trade-union membership. The many dangers of personal data in the wrong hands are well known: it can be used for blackmail, coercion, social-engineering scams, personality theft, or sold to advertising and insurance companies. Live personal data can even be used by organised criminals who want to commit theft or kidnapping or by terrorist organisations for targetted attacks.

Personal information is not the only type of information that can be sensitive. Information about critical infrastructures is also likely to be collected and recombined as part of emergency computation. Such information can be valuable to criminals, potential terrorists, and foreign powers. Although many of the data sets used will be in the open, their potential value for adversaries increase when they are recombined and augmented with temporal information that can be used, for example, by criminals or terrorists to estimate police and other response times.

Finding the right balance between emergency needs and ethical/societal concerns requires an appropriate combination of organisational, informational, and technical measures. On the \textit{organisational side}, sensitive data should only be made accessible to trusted agencies with clear procedures in place for screening personnel etc. For every data set that includes sensitive information, a steward should be appointed with clearly-defined responsibilities. As much as possible, data should be collected only from trusted sources, but this is not always an option --- and certainly not in the case of socially generated data. And, of course, data should only be shared with trusted partners.

On the \textit{informational side}, sensitive data should only be collected and recombined in response to concrete and carefully prioritised operational and tactical needs, even when it goes against the opportunistic tendency of big-data practice, which stores data opportunistically --- exhaustively, historically, and in full detail --- because it can potentially become valuable in the future. Data tables should be projected to remove unneeded attributes whenever they are shared. Whenever possible, anonymisation and pseudonymisation should be used to increase privacy, although they cannot be considered sufficient privacy measures in their own right.

On the \textit{technical side}, only screened and trusted data processing organisations should be allowed to store and process the data. This applies in particular to cloud computing providers. The processing organisations should be transparent when it comes to where in the world --- and thus under which jurisdiction --- the data is stored and processed. Techniques such as secret sharing and watermarking should be considered to make the data more difficult to obtain and leave it traceable should it come into the wrong hands.

\section*{Exercises}
\begin{enumerate}
  \item  What are the three V's of big data? 
  \item  Some people talk about two more V's in addition to the three. Which ones?
  \item  What does it mean that big data are: exhaustive, fine grained, indexical, relational, extendible, historical, and opportunistic.
  \item  What is a data-driven organisation (or business)? Give examples?
  \item  Explain the four phases of emergency management. How can big data help in each of them?
  \item  Which big data sources can be leveraged in each phase of emergency management and how?
  \item  Name the most central big-data technologies for: file management, database management, and data processing.
  \item  What are the main improvements of Spark over Google's MapReduce and Apache Hadoop?
  \item  What is an RDD and what is lineage in Spark?
  \item  What are the most pressing ethical and societal dangers of big data?
  \item  In the Spark example, change the code to output Named Entities (names of individual things like people, places, organisations, and works) instead of hashtags, assuming that named entities are always written as a sequence of words with capital initial letters.
\end{enumerate}

%
\bibliographystyle{spbasic}              	
\bibliography{References-BDEMTextbook}	

\section*{Acknowledgement}

This is a pre-print of the following chapter: Opdahl, A. L. and Nunavath, V., ``Big Data'', published in Big Data in Emergency Management: Exploitation Techniques for Social and Mobile Data, edited
by Rajendra Akerkar, 2020, Springer International Publishing reproduced with
permission of Springer International Publishing. The final authenticated version is
available online at: \url{http://dx.doi.org/10.1007/978-3-030-48099-8}.

\mbox{}

\noindent
Opdahl, A. L., and Nunavath, V. (2020). Big Data. Big Data in Emergency Management: Exploitation Techniques for Social and Mobile Data, 15-29.

\end{document}